\documentclass[twocolumn,showpacs,preprintnumbers,amsmath,amssymb]{revtex4}
\usepackage{graphicx}% Include figure files
\usepackage{dcolumn}% Align table columns on decimal point
\usepackage{bm}% bold math

\begin{document}

\setlength{\topmargin}{0pt}
\setlength{\parskip}{0.5\baselineskip}

\preprint{/}

\title{Electron correlation in the two-dimensional triangle lattice of Na$_x$CoO$_2$ 
}% Force line breaks with \\

\author{J. Sugiyama$^1$}
 \email{e0589@mosk.tytlabs.co.jp}
\author{J. H. Brewer$^2$}
\author{E. J. Ansaldo$^3$}
\author{B. Hitti$^3$}%
\author{M. Mikami$^4$}%
 \altaffiliation[Present address: ]{National Institute of Advanced Industrial Science and technology, Ikeda, Osaka 563-8577, Japan.}
\author{Y. Mori$^4$}%
\author{T. Sasaki$^4$}%
\affiliation{%
$^1$Toyota Central Research and Development Labs. Inc., 
 Nagakute, Aichi 480-1192, Japan}%

\affiliation{
$^2$TRIUMF, CIAR and 
Department of Physics and Astronomy, University of British Columbia, 
Vancouver, BC, V6T 1Z1 Canada 
% with \\
}%

\affiliation{
$^3$TRIUMF, 4004 Wesbrook Mall, Vancouver, BC, V6T 2A3 Canada   
% with \\
}%

\affiliation{
$^4$Department of Electrical Engineering, Osaka University, Suita, Osaka 563-8577, Japan   
% with \\
}%

\date{\today}% \today is always today, 
             %  but any date may be explicitly specified

\begin{abstract}
Magnetism of layered cobaltites 
Na$_x$CoO$_2$ with $x$ = 0.6 and 0.9
has been investigated by 
a positive muon spin rotation and relaxation 
($\mu^+$SR) spectroscopy
together with 
magnetic susceptibility and specific heat 
measurements, 
using single crystal samples in the
temperature range between 250 and 1.8~K. 
Zero-field (ZF-) $\mu^+$SR measurements on 
Na$_{0.9}$CoO$_2$ indicates a transition from a paramagnetic to 
an incommensurate spin density wave state at 19~K(=$T_{\sf SDW}$).
The anisotropic ZF-$\mu^+$SR spectra suggest that 
the oscillating moments of the {\sf IC-SDW} directs 
along the $c$-axis. 
Since Na$_{0.6}$CoO$_2$ is paramagnetic down to 1.8~K, 
the magnitude of $T_{\sf SDW}$ is found to strongly depend on $x$.
This behavior is well explained using the Hubbard model 
within a mean field approximation 
on two-dimensional triangle lattice 
in the CoO$_2$ plane.
Also, both the appearance of the {\sf IC-SDW} state by the change in $x$ 
and the magnitude of the electronic specific heat parameter of 
Na$_{0.6}$CoO$_2$
indicate that 
Na$_x$CoO$_2$ is unlikely to be 
a typical strongly correlated electron system. 
\end{abstract}

\pacs{76.75.+i, 75.30.Fv, 72.15.Jf, 65.40.Ba}% 
% (PACS, the Physics and Astronomy Classification Scheme)
\keywords{Thermoelectric layered cobaltites, magnetism, 
 specific heat, muon spin rotation, incommensurate spin density waves}
% (Use showkeys class option if keyword display desired)

\maketitle

\section{\label{sec:Intro}Introduction}
%:\protect\\\textbackslash\textbackslash}

Layered cobaltites with a two-dimensional-triangular lattice ({\sf 2DTL})
of Co ions have been intensely investigated because of 
their structural variety, 
promising thermoelectric performance, 
\cite{NCO_1,NCO_2,NCO_3,CCO_1,CCO_2,CCO_3,4LBiSCO_1,4LCCCO_1}
and unpredictable superconducting behavior 
induced by an intercalation of H$_2$O. 
\cite{NCOsc_1,NCOsc_2,NCOsc_3}
Among them, the sodium cobaltite Na$_x$CoO$_2$ is apparently  
the most basic compound, because of its simple structure,
the single CoO$_2$ planes and the
single disordered Na planes 
form alternating stacks along the $c$ axis.\cite{NCO_structure_1}
The CoO$_2$ planes, in which the {\sf 2DTL} of Co 
ions is formed by a network of edge-sharing CoO$_6$
octahedra, are the conduction planes of Na$_x$CoO$_2$.
Therefore, the magnetic and transport properties of 
Na$_x$CoO$_2$ are expected to strongly depend on 
the nominal valence of the Co ions in the CoO$_2$ planes, 
similar to the case of the CuO$_2$ planes for the high-$T_c$ cuprates.

Motohashi{\it et al.} reported the existence of a
magnetic transition at 22~K (= $T_{\rm m}$) in 
polycrystalline Na$_{0.75}$CoO$_2$ 
from the observation of small changes in bulk susceptibility 
and transport properties, 
while no transitions were found 
in Na$_{0.65}$CoO$_2$ down to 2~K.\cite{NCO_TIT2} 
Recent positive muon spin rotation and relaxation ($\mu^+$SR) 
experiment on a polycrystalline Na$_{0.75}$CoO$_2$ sample 
\cite{jun_PRB4} 
indicated that the transition at 22~K 
is not induced by impurities 
but is an intrinsic change in the magnetism
of the sample, 
although the sample was 
magnetically inhomogeneous. 
Furthermore, the $\mu^+$SR result suggested that 
the ordered phase below $T_{\rm m}$ could be either 
a ferrimagnetic ({\sf FR}) or 
a commensurate ({\sf C}) spin density wave ({\sf SDW}) state.\cite{jun_PRB4}

On the other hand, the related compound, 
pure and doped 
[Ca$_2$CoO$_3$]$_{0.62}^{\rm RS}$[CoO$_2$] 
(RS denotes the rocksalt-type susbsystem),
exhibited a transition to 
an incommensurate ({\sf IC}) {\sf SDW} state 
below $\sim$100~K.\cite{jun_PRB1}  
Recent $\mu^+$SR experiment using both single crystals and 
$c$-aligned polycrystals suggested 
that long-range {\sf IC-SDW} order forms 
below $\sim$ 30~K ($\equiv T_{\sf SDW}$), 
while a short-range order appears 
below 100~K ($\equiv T_{\rm c}^{\rm on}$).
\cite{jun_PRB2, jun_PRB3} 
Also, the {\sf IC-SDW} was found to be induced 
by the ordering of the Co spins 
in the CoO$_2$ planes.

The magnetically ordered state is, therefore, most likely common 
for the {\sf 2DTL} of the Co ions in the layered cobaltites.
The relationship between the transition temperature and 
the Co valence is, therefore, worth to investigate 
in order to better understand the nature of the {\sf 2DTL}.
In particular, the Co valence in the CoO$_2$ planes 
changes directly in proportion to $x$ for Na$_x$CoO$_2$,
whereas that is unclear for
[Ca$_2$CoO$_3$]$_{0.62}^{\rm RS}$[CoO$_2$] 
because of the two unequivallent Co sites in the lattice.
It should be noted that the polycrystalline Na$_{0.75}$CoO$_2$ sample,  
although structurally single phase, 
was found to be magnetically inhomogeneous.\cite{jun_PRB4}
Hence, the $\mu^+$SR experiments 
on single crystals are important 
for further elucidation of the nature of the {\sf 2DTL}.

The detailed crystal structure of Na$_x$CoO$_2$ 
was reported to depend on $x$ and reaction temperature;
\cite{NCO_structure_1} 
that is, $\alpha$-Na$_x$CoO$_2$ with 0.9$\leq x\leq$1, 
$\alpha$'-Na$_x$CoO$_2$ with $x$=0.75, 
$\beta$-Na$_x$CoO$_2$ with 0.55$\leq x\leq$0.6 and 
$\gamma$-Na$_x$CoO$_2$ with 0.55$\leq x\leq$0.74. 
Also, the structure of Na$_{0.5}$CoO$_2$ 
was reported to be assigned as the $\gamma$-phase.
\cite{NCO_structure_2} 
Unfortunately, single crystals are available only for 
the $\alpha$-Na$_x$CoO$_2$ phase with $x$=0.9 and 
the $\gamma$-Na$_x$CoO$_2$ phase with $x$=0.6,\cite{NCO_Mika1} 
and attempts to prepare crystals with the other $x$ 
have failed so far. 
Magnetic susceptibility ($\chi$) measurements 
on these crystals indicated that 
$\alpha$-Na$_{0.9}$CoO$_2$ exhibits 
a magnetic transition at $\sim$20~K, 
whereas $\gamma$-Na$_{0.6}$CoO$_2$ seems 
to be a Curie-Weiss paramagnet down to 5~K.\cite{NCO_Mika2} 
Also, the resistivity ($\rho$)-vas-$T$ curve in 
$\alpha$-Na$_{0.9}$CoO$_2$ 
was found to be metallic above $\sim$20~K, 
while semiconducting below $\sim$20~K.\cite{NCO_Mika1,NCO_Mika2} 

We report both weak (65~Oe) 
transverse-field (wTF-) $\mu^+$SR 
and zero field (ZF-) $\mu^+$SR measurements 
for single crystal platelets of 
Na$_{0.9}$CoO$_2$ and 
Na$_{0.6}$CoO$_2$
at temperatures below 300~K 
to further elucidate the relationship between 
the transition temperature and the Co valence, 
{\it i.e.}, the magnetic phase diagram.
In addition, we performed heat capacity measurements 
on these crystals 
to study the magnetism of Na$_x$CoO$_2$ in full detail.

\section{\label{sec:Expt}Experimental}
%:\protect\\\textbackslash\textbackslash}

Single crystals of Na$_x$CoO$_2$ were grown 
at Osaka University by a flux technique
using reagent grade Na$_2$CO$_3$ 
and Co$_3$O$_4$ powders as starting materials.
A mixture of NaCl, Na$_2$CO$_3$ and B$_2$O$_3$ 
was used as the flux.
The typical dimension of the obtained Na$_{0.9}$CoO$_2$ platelets
was $\sim 3\times3\times$0.1~mm$^3$,
while that of Na$_{0.6}$CoO$_2$ was
$\sim 6\times6\times$0.1~mm$^3$.
The preparation and characterization of these crystals 
were reported in greater detail elsewhere.\cite{NCO_Mika1,NCO_Mika2} 

Susceptibility ($\chi$) was measured 
using a superconducting quantum interference device 
(SQUID) magnetometer 
(mpms, Quantum Design) 
in the temperature range between 400 and 5~K 
under magnetic field $H \leq$ 55~kOe. 
In order to increase the magnetic signal, 
5 platelets were stacked in a plastic sample holder. 
Also, to determine anisotropy, 
$H$ was applied parallel or perpendicular to the basal ({\it i.e., c}) plane. 
Hereby, we will abbreviate susceptibility obtained with 
$H //  c$ as $\chi_c$ and 
$H \bot c$ as $\chi_a$, respectively.

Heat capacity ($C_{\rm p}$) was measured 
using a relaxation technique
(ppms, Quantum Design) 
in the temperature range between 
300 and 1.9~K. 
The $\mu^+$SR experiments were performed on 
the {\sf M20} and {\sf M15} surface muon beam lines at TRIUMF. 
The experimental setup and techniques
were described elsewhere.\cite{ICSDW_1} 

\section{\label{sec:Results}Results}
\subsection{\label{ssec:SH} susceptibility and heat capacity}
\begin{figure}
\includegraphics[width=8cm]{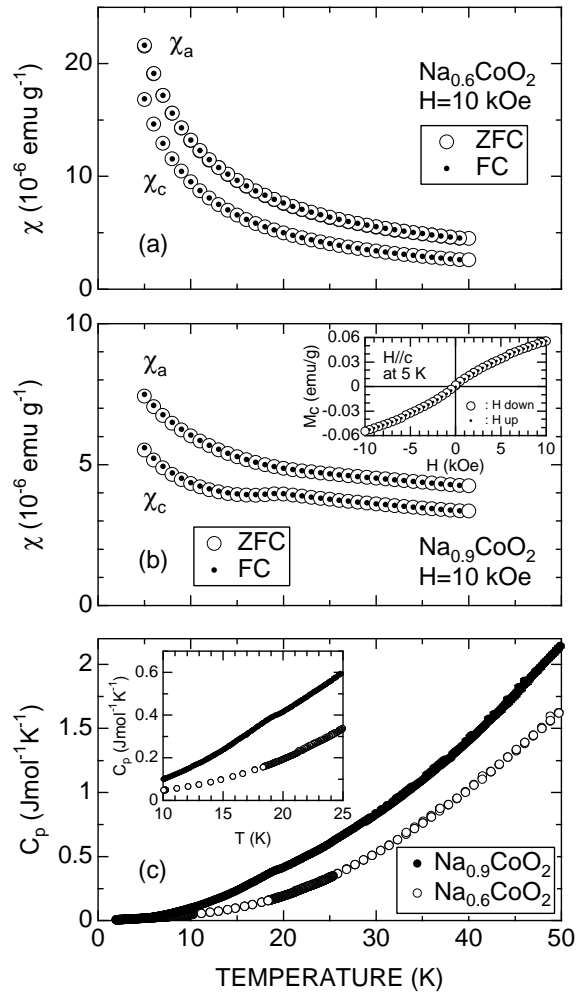}
\caption{\label{fig:Cp}(a) Temperature dependences of 
magnetic susceptibility $\chi$ for Na$_{0.6}$CoO$_2$ and 
(b) Na$_{0.9}$CoO$_2$ and 
(c) temperature dependence of heat capacity 
$C_{\rm p}$ for Na$_{0.9}$CoO$_2$ and Na$_{0.6}$CoO$_2$. 
The inset in Fig.~\ref{fig:Cp}(b) shows 
the magnetization $M$ as a function of magnetic field $H$ at 5~K 
and that in Fig.~\ref{fig:Cp}(c) the magnification of 
the $C_{\rm p}(T)$ curve in the vicinity of $T_{\rm c}$.
$\chi$ was measured both in zero field-cooling $ZFC$ 
and field-cooling $FC$ mode with $H=10$~kOe and 
$H$ applied parallel and perpendicular to the $c$ axis.
}
\end{figure}
\begin{figure}
\includegraphics[width=8cm]{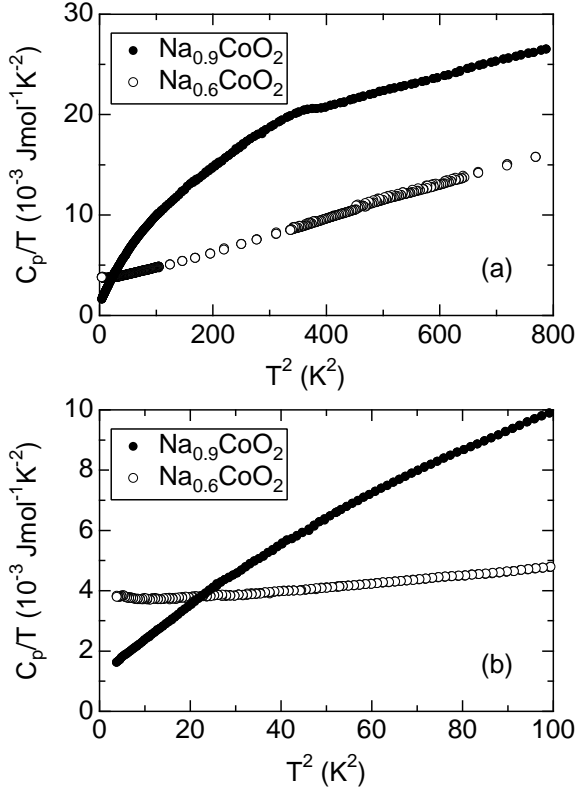}
\caption{\label{fig:Cp2}(a) The relationship between  
$C_{\rm p}/T$ and $T^2$ 
for Na$_{0.9}$CoO$_2$ and Na$_{0.6}$CoO$_2$; 
(a) 0$\leq T^2\leq$800~K$^2$ and (b) 0$\leq T^2\leq$100~K$^2$
}
\end{figure}

The temperature dependence of $\chi_c$ 
for the Na$_{0.9}$CoO$_2$ crystal exhibits 
a clear cusp at 19~K (=$T_{\rm c}$), 
whereas no clear anomalies in the $\chi_a(T)$ curve
(see Fig.~\ref{fig:Cp}(b)).
Since there is no marked difference between 
the data obtained in zero field cooling ($ZFC$) 
and field cooling ($FC$) mode, 
this cusp is most likely due to 
an antiferromagnetic transition.
Actually, the relationship between magnetization $M_c$ and $H$ 
has no loop even at 5~K (see the inset of Fig.~\ref{fig:Cp}(b))
The $\chi(T)$ curves for the Na$_{0.6}$CoO$_2$ crystal show 
a paramagnetic behavior down to 5~K with a small anisotropy, 
as reported.\cite{NCO_Mika2} 

The $C_{\rm p}(T)$ curve for the Na$_{0.9}$CoO$_2$ crystal 
also exhibits a small maximum at 19~K, 
indicating the transition detected by the $\chi$ measurement. 
On the other hand, $C_{\rm p}$
for the Na$_{0.6}$CoO$_2$ crystal decreases monotonically 
with decreasing $T$ down to 1.9~K. 
Note that in Fig.~\ref{fig:Cp}(c) 1~mol denotes 1~mol atom; thus, 
$C_{\rm p}$ for Na$_x$CoO$_2$ is equivalent to 
the measured heat capacity (Jg$^{-1}$K$^{-1}$) 
divided by (3+$x$)$M$, 
where $M$ is the molar weight of Na$_x$CoO$_2$.
The dependence of $C_{\rm p}/T$ on $T^2$ 
for Na$_{0.9}$CoO$_2$ looks very complicated 
due to a contribution of the magnetic order, 
although the $C_{\rm p}/T$-vas-$T^2$ curve 
for Na$_{0.6}$CoO$_2$ exhibits an almost linear relation 
in the $T^2$ range above 30~K$^2$ 
(see Fig.~\ref{fig:Cp2}). 
This is also evidence for lack of magnetic order 
in Na$_{0.6}$CoO$_2$ down to 1.9~K. 

If we employ the Debye formula
\begin{eqnarray}
\frac{C_{\rm p}}{T} &=& \gamma + \beta T^2 ,
\label{eq:Debye}
\end{eqnarray}
in the $T^2$ range between 30 and 100~K$^2$,
we obtain the electronic specific heat parameter 
$\gamma$ = 3.62$\pm$0.04~mJK$^{-2}$ per mol atom and 
$\beta$ = 0.0614$\pm$0.0004~mJmolK$^{-4}$ per mol atom. 
This gives the Debye temperature ($\theta _{\rm D}$) 
of 316~K. 
Two different platelets of Na$_{0.6}$CoO$_2$ provided 
almost same values of $\gamma$ and $\theta _{\rm D}$.
Making comparison with the data 
for the polycrystalline Na$_{0.55}$CoO$_2$ 
($\gamma$ = 6.8~mJK$^{-2}$ per mol atom and 
$\theta _{\rm D}$=354~K),\cite{NCO_5}  
the present values are rather small, 
probably due to the effects of 
grain boundaries and/or undetected second phases 
in the polycrystals. 
Furthermore, the present $\gamma$ is 
in good agreement with the value calculated 
for band structure of 
Na$_{0.5}$CoO$_2$ 
($\gamma$ = 3~mJK$^{-2}$ per mol atom).\cite{NCO_6} 

\subsection{\label{ssec:wTF} wTF-$\mu^+$SR}

\begin{figure}
\includegraphics[width=8cm]{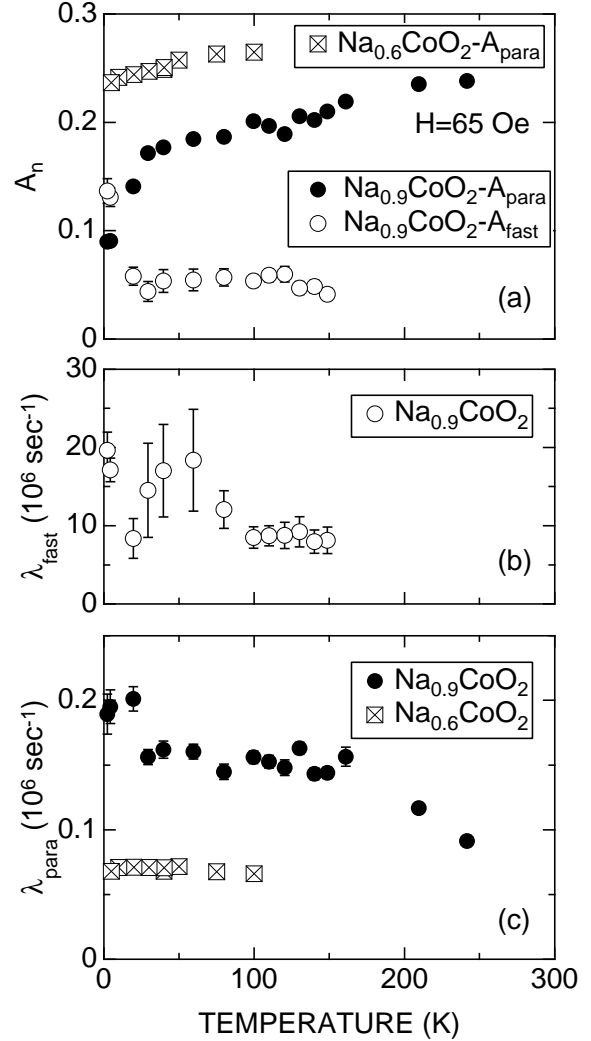}
\caption{\label{fig:wTF-muSR}Temperature dependences of 
(a) $A_{\sf para}$ and $A_{\sf fast}$, 
(b) $\lambda_{\sf fast}$ and 
(c) $\lambda_{\sf para}$ in single crystal platelets of 
Na$_{0.9}$CoO$_2$ (solid and open circles) 
and Na$_{0.6}$CoO$_2$ (crossed squares). 
The data were obtained by fitting 
the wTF-$\mu^+$SR spectra with Eq. (\ref{eq:TFfit}).
}
\end{figure}

The wTF-$\mu^+$SR spectra in a magnetic field of 
$H\sim 65$~Oe in the 
Na$_{0.9}$CoO$_2$ crystals 
exhibit a clear reduction of the $\mu^+$ precession 
paramagnetic amplitude below $T_{\rm c}$.  
The wTF-$\mu^+$SR spectrum below $T_{\rm c}$ was 
well fitted 
in the time domain with a combination of 
a slowly relaxing precessing signal and 
a fast non-oscillatory signal, as in the case of 
[Ca$_2$CoO$_3$]$_{0.62}^{\rm RS}$[CoO$_2$]:\cite{jun_PRB1}
\begin{eqnarray}
A_0 \, P(t) &=& A_{\sf para} \, \exp(- \lambda_{\sf para} t) \, \cos (\omega_\mu t + \phi)
\cr
&+& A_{\sf fast} \, \exp(-\lambda_{\sf fast} t) ,
\label{eq:TFfit}
\end{eqnarray}
where $A_0$ is the initial asymmetry, 
$P(t)$ is the muon spin polarization function, 
$\omega_\mu$ is the muon Larmor frequency, 
$\phi$ is the initial phase of the precession and 
$A_n$ and $\lambda_n$ ($n$ = {\sf para} and {\sf fast}) 
are the asymmetries and exponential relaxation rates of the two signals.  
Interestingly, the non-oscillatory signal ($n$ = {\sf fast}) 
has finite amplitudes below $\sim 150$~K.

Figures~\ref{fig:wTF-muSR}(a) - \ref{fig:wTF-muSR}(c) 
show the temperature dependences of 
$A_{n}$ and $\lambda_{n}$ ($n$ = {\sf para} and {\sf fast})
in Na$_{0.9}$CoO$_2$ and Na$_{0.6}$CoO$_2$. 
For Na$_{0.9}$CoO$_2$, as $T$ decreases from 250~K, 
the magnitude of $A_{\sf para}$ 
decreases monotonically down to $T_{\rm c}$, 
then slope (d$A_{\sf para}$/d$T$) suddenly steepens 
with further lowering $T$.  
Also, the $A_{\sf fast}(T)$ curve 
is nearly constant down to $T_{\rm c}$ 
and then suddenly increases with $T$.
On the other hand, as $T$ decreases from 250~K, 
$\lambda_{\sf para}(T)$ increases down to 150~K, 
roughly levels off to a constant value 
down to $T_{\rm c}$, 
and then suddenly increases at $T_{\rm c}$ and 
again levels off to a constant value below $T_{\rm c}$. 
Considering the large experimental error, 
the $\lambda_{\sf fast}(T)$ curve is approximately 
a constant down to $T_{\rm c}$ 
and then suddenly increases with $T$.

These behaviors clearly indicate the existence of 
a transition to a magnetically ordered state 
at $T_{\rm c}$ in Na$_{0.9}$CoO$_2$, 
as expected by the $\chi$ and $C_{\rm p}$ measurements. 
Furthermore, the $\mu^+$SR data show other changes 
at around 150~K.
The transport measurements 
(resistivity $\rho$ and Seebeck coefficient $S$),
both the d$\rho$/d$T(T)$ and $S(T)$ curves 
exhibited a broad maximum around 150~K,
whereas no clear anomalies were seen 
in the $\chi(T)$ curve.\cite{NCO_Mika1,NCO_Mika2}

Furthermore, there are no marked changes in both 
$A_{\sf para}(T)$ and $\lambda_{\sf para}(T)$ 
in Na$_{0.6}$CoO$_2$. 
This indicates that the Na-poorer crystal 
is a paramagnet down to 5~K,
consistent with the results of the $\chi$ 
and $C_{\rm p}$ measurements.

\subsection{\label{ssec:ZF} ZF-$\mu^+$SR}
\begin{figure}
\includegraphics[width=8cm]{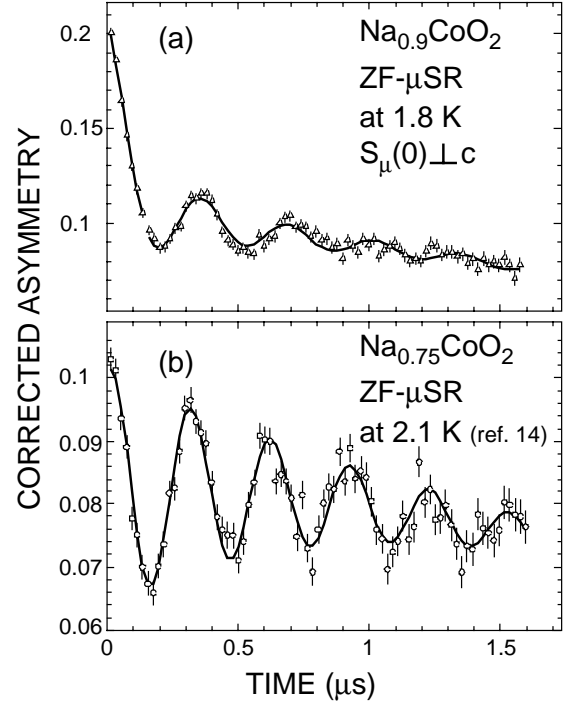}
\caption{\label{fig:ZF-muSR}  ZF-$\mu^+$SR time spectra of 
  (a) single crystal platelets of Na$_{0.9}$CoO$_2$ at 1.8~K and 
  (b) a polycrystalline plate of Na$_{0.75}$CoO$_2$ at 2.1~K (ref. ~14).
  In Fig.~\ref{fig:ZF-muSR}(a), 
  the configuration of the sample and the initial muon spin direction 
  $\vec{\bm S}_\mu(0)$ is $\vec{\bm S}_\mu(0) \perp \hat{\bm c}$. 
  Both a rapid initial decay of the amplitude 
  and a delay of the initial phase in the top spectrum 
  indicate that an incommensurate spin structure presents in Na$_{0.9}$CoO$_2$, 
  whereas a commensurate structure in Na$_{0.75}$CoO$_2$.
}
\end{figure}
Figure \ref{fig:ZF-muSR}(a) shows ZF-$\mu^+$SR time spectrum at 1.8~K 
in the single crystal platelets of Na$_{0.9}$CoO$_2$; 
the spectrum was obtained with 
the initial $\mu^+$ spin direction $\vec{\bm S}_\mu(0)$ 
perpendicular to the $c$-axis.  
A clear oscillation due to quasi-static internal fields 
is observed only when $\vec{\bm S}_\mu(0) \perp \hat{\bm c}$.
Also, the ZF-$\mu^+$SR time spectrum 
for the polycrystalline Na$_{0.75}$CoO$_2$ sample, 
which entered a commensurate spin structure below 22~K, \cite{jun_PRB4}
is shown in Fig.~\ref{fig:ZF-muSR}(b). 
Making comparison with the bottom spectrum, 
the oscillation amplitude in the top spectrum decays rapidly and 
the initial phase delays.   
The oscillation in the top spectrum is characteristic of 
a zeroth-order Bessel function of the first kind $J_0(\omega_\mu t)$ 
that describes the muon polarization evolution 
in an {\sf IC-SDW} field distribution.\cite{ICSDW_2,ICSDW_3} 
Actually, the top oscillating spectrum was fitted
using a combination of three signals: 
\begin{eqnarray}
 A_0 \, P(t) &=& 
   A_{\sf SDW} \, J_0(\omega_{\mu} t) \, \exp(-\lambda_{\sf SDW} t)^{\beta_{\sf SDW}}
\cr
 &+& A_{\rm KT} \, G_{zz}^{\rm KT}(t,\Delta) 
\cr
 &+& A_{\rm x} \, \exp(-\lambda_{\rm_x} t)^{\beta_{\rm x}},
\label{eq:ZFfit}
\end{eqnarray}
\begin{equation}
 \omega_\mu \equiv  2 \pi \nu_\mu = \gamma_{\mu} \; H_{\sf int},
\label{eq:omg}
\end{equation}
\begin{eqnarray}
 G_{zz}^{\rm KT}(t,\Delta) &=&
 \frac{1}{3} + \frac{2}{3} \, \left( 1 - \Delta^2 t^2 \right) 
 \exp(- \frac{\Delta^2 t^2} { 2}),
\label{eq:GKT}
\end{eqnarray}
where $A_0$ is the empirical maximum muon decay asymmetry, 
$A_{\sf SDW}$, $A_{\rm KT}$ and $A_{\rm x}$ 
are the asymmetries associated with the three signals, 
$G_{zz}^{\rm KT}(t,\Delta)$ is the static Gaussian Kubo-Toyabe function, 
$\Delta$ is the static width of the distribution of local frequencies 
at the disordered sites,
$\lambda_{\rm x}$ is the slow relaxation rate and
$\beta_{\sf SDW}$ and $\beta_{\rm x}$ are 
the power of the exponential relaxation. 
Fits using just an exponentially damped cosine oscillation,
$\exp(-\lambda t) \cos(\omega_{\mu} t + \phi)$,
provides a phase angle $\phi \sim -42^{\rm o}$, 
which is physically meaningless,\cite{ICSDW_4}
although $\phi \sim 0^{\rm o}$ for the bottom spectrum. \cite{jun_PRB4}

We therefore conclude that Na$_{0.9}$CoO$_2$ undergoes 
a magnetic transition from a paramagnetic state 
to an {\sf IC-SDW} state 
({\it i.e.} $T_c = T_{\sf SDW}$).  
The absence of a clear oscillation in the spectrum 
obtained with $\vec{\bm S}_\mu(0) \parallel \hat{\bm c}$ indicates that 
the internal magnetic field $\vec{\bm H}_{\rm int}$ 
is roughly parallel to the $c$-axis, since the muon spins 
do not precess in a parallel magnetic field.  
The {\sf IC-SDW} is unlikely to propagate along the $c$-axis 
due to the two-dimensionality.
The {\sf IC-SDW} is therefore concluded 
to propagate in the $c$ plane ({\it i.e.}, the [CoO$_2$] plane), 
with oscillating moments directed along the $c$-axis, 
as in the case of 
[Ca$_2$CoO$_3$]$_{0.62}^{\rm RS}$[CoO$_2$].\cite{jun_PRB3} 
Moreover, this anisotropic result is consistent with 
$\chi$ measurements. 
That is, the cusp at 19~K was observed, 
when $H$ was applied parallel to the $c$ axis, 
whereas the cusp was undetected if $H\perp c$.
\cite{NCO_Mika2} 

\begin{figure}
\includegraphics[width=8cm]{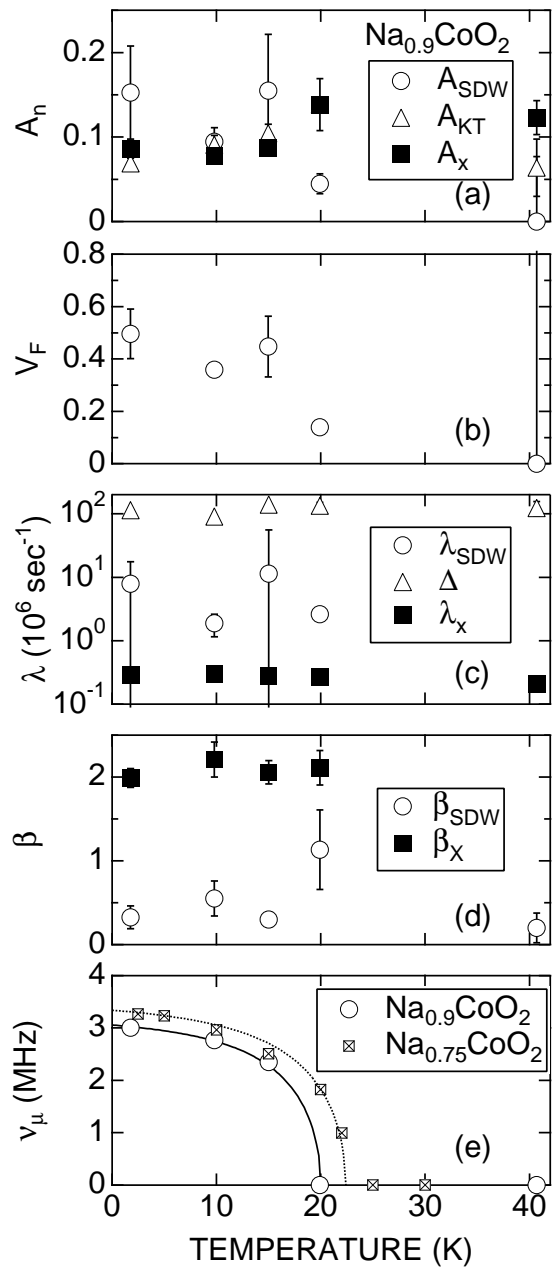}
\caption{\label{fig:ZFmuSR2} 
 Temperature dependences of 
(a) $A_{\sf SDW}$, $A_{\sf KT}$ and $A_{\sf x}$, 
(b) the volume fraction of the {\sf SDW} signal ($V_{\rm F}$), 
(c) $\lambda_{\sf SDW}$, $\Delta$ and $\lambda_{\sf x}$ 
(d) $\beta _{\sf SDW}$ and $\beta _{\sf x}$
(e) $\nu_{\mu}$ 
 for the single crystal platelets of 
 Na$_{0.9}$CoO$_2$. 
 In Fig.~\ref{fig:ZFmuSR2}(e), the $\nu_{\mu}(T)$ curve 
 for the polycrystalline Na$_{0.75}$CoO$_2$ sample (ref.~14) 
 is also shown, 
 although its ZF-$\mu^+$SR spectrum 
 was fitted by an exponentially damped cosine oscillation, 
 $\exp(-\lambda t) \cos(\omega_{\mu} t + \phi)$. 
 The solid and broken lines in Fig.~\ref{fig:ZFmuSR2}(e) represent 
 the temperature dependence of the {\sf BCS} gap energy. 
 }
\end{figure}
Figures~\ref{fig:ZFmuSR2}(a)-\ref{fig:ZFmuSR2}(e) show
the temperature dependences of 
(a) $A_{\sf SDW}$, $A_{\sf KT}$ and $A_{\sf x}$, 
(b) the volume fraction of the {\sf SDW} signal ($V_{\rm F}$), 
(c) $\lambda_{\sf SDW}$, $\Delta$ and $\lambda_{\sf x}$ and 
(d) $\beta _{\sf SDW}$ and $\beta _{\sf x}$
(e) $\nu_{\mu}$ 
for the single crystal platelets of Na$_{0.9}$CoO$_2$.
The volume fraction ($V_{\rm F}$) was calculated as 
$A_{\sf SDW}$/($A_{\sf SDW}+A_{\rm KT}+A_{\sf x}$).
The magnitude of $A_{\sf SDW}$ increases 
with decreasing $T$ below $T_{\sf SDW}$, 
although both $A_{\rm KT}$ and $A_{\rm x}$ 
are almost $T$ independent. 
(see Fig.~\ref{fig:ZFmuSR2}(a)). 
Here, the static Gaussian Kubo-Toyabe function 
is for the signal from muon sites experiencing 
disordered magnetic fields, and 
the the power exponential relaxation 
is due to fast fluctuating magnetic fields.
\cite{ICSDW_2,ICSDW_3} 
Thus, only the $A_{\sf SDW}$ signal 
appears below $T_{\sf SDW}$,
while the latter two signals are not affected 
by the formation of the {\sf IC-SDW} order. 
Actually, the other parameters of the latter two signals, 
{\it i.e.}, $\lambda_{\sf x}$, $\Delta$ and $\beta _{\sf x}$, 
also seem to be independent of $T$. 
Hereby, we, therefore, ignore the contributions 
from these two signals.

The $V_{\rm F}(T)$ curve increases monotonically 
with decreasing $T$ below $T_{\sf SDW}$, 
reaching a maximum at the lowest $T$ 
is $V_{\rm F}$=50\% at 1.8~K. 
This value is significantly higher than 
in polycrystalline 
Na$_{0.75}$CoO$_2$ ($V_{\rm F}$=21\%).\cite{jun_PRB4} 
However, $V_{\rm F}$ for the $c$-aligned 
[Ca$_2$CoO$_3$]$_{0.62}^{\rm RS}$[CoO$_2$] and
[Ca$_2$Co$_{4/3}$Cu$_{2/3}$O$_4$]$_{0.62}^{\rm RS}$[CoO$_2$]
samples were found to be $\sim$100\%.\cite{jun_PRB3} 
Thus, the present Na$_{0.9}$CoO$_2$ crystals are considered to 
be still magnetically inhomogeneous,
probably due to an inhomogeneous distribution of the Na ions. 
In particular, Na$_{0.9}$CoO$_2$ 
is reported to be unstable in humid air;\cite{NCO_Mika2} 
the excess Na easily reacts with moisture in air,
although the present samples were kept in a desiccator 
prior to the $\mu^+$SR experiment. 
Both the temperature dependences of 
$\lambda_{\sf SDW}$ and $\beta_{\sf SDW}$ are fairy flat 
(see Figs.~\ref{fig:ZFmuSR2}(c) and (d)),
while, as $T$ decreases, the $\nu_{\mu}$ increases 
with a decreasing slope $d\nu_{\mu}/dT$ 
(Fig.~\ref{fig:ZFmuSR2}(e)). 
This behavior is well explained by 
the {\sf BCS} prediction for order parameter 
for the {\sf SDW} state.\cite{ICSDW_5} 
Since the value of $\nu_{\mu}$(0~K) is estimated as 3.0~MHz, 
the internal magnetic field in the magnetically ordered state 
in Na$_{0.9}$CoO$_2$ is almost equivalent to that in the polycrystalline 
Na$_{0.75}$CoO$_2$ sample ($\nu_{\mu}$(0~K)$\sim$3.3~MHz).\cite{jun_PRB4} 
\section{\label{sec:Discussion}Discussion}
\subsection{\label{ssec:PE} phase diagram and electron correlation}
\begin{figure}
\includegraphics[width=8cm]{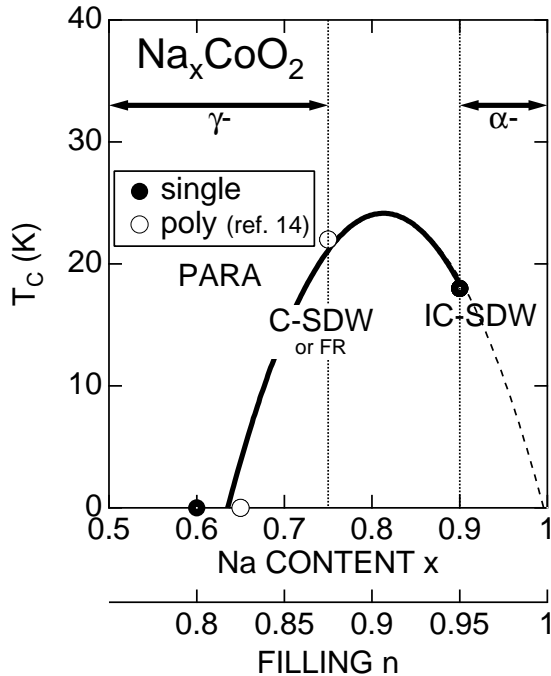}
\caption{\label{fig:Phase} Phase diagram of Na$_x$CoO$_2$ 
determined by $\mu^+$SR experiments. Solid circles represent 
the present result and open circles the previous data (ref.~14). 
The point at $x$=1 is a speculation from 
the data on the related compound LiCoO$_2$.
The relationship between structure and $x$ is also shown;
$\alpha$ and $\gamma$ denote the $\alpha$-phase and $\gamma$-phase.
The $\beta$-phase (0.55$\leq x\leq$0.6) is unstable at high temperatures 
used for the sample preparation in the current work and changes 
to the $\gamma$-phase;
hence, only the two ($\alpha$ and $\gamma$) phases are presented.
}
\end{figure}
In the previous $\mu^+$SR experiment 
\cite{jun_PRB4} 
on polycrystalline 
Na$_{0.75}$CoO$_2$ and Na$_{0.65}$CoO$_2$ samples, 
the Na-richer sample exhibited a transition to either 
a ferrimagnetic ({\sf FR}) or 
a commensurate ({\sf C-}) {\sf SDW} state at 22~K, 
whereas the Na-poorer sample was a paramagnet down to 2~K. 
Since the two samples were single phase of 
a hexagonal structure ($\gamma$-phase),\cite{NCO_TIT1}  
the magnetic transition at 22~K is considered to be 
an intrinsic behavior in Na$_{0.75}$CoO$_2$. 
Therefore, we can sketch the magnetic phase diagram of 
Na$_x$CoO$_2$ with $x\geq$0.6 
as a function of $x$
(see Fig.~\ref{fig:Phase}).
Recent compositional and chemical titration analyses 
indicated that the oxygen deficiency $\delta$ in 
Na$_x$CoO$_{2-\delta}$ 
is negligibly small even for the sample with $x$=0.9.
\cite{NCO_Mika2} 
This means that the average Co valence is directly calculated from $x$.
As $x$ increases from 0.6 ({\it i.e.}, the Co valence decreases from 3.4), 
the magnitude of $T_{\rm C}$ increases 
up to a maximum at around $x$=0.8, 
decreases with further increasing $x$. 
As a result, we obtain a dome-shaped relationship between 
$T_{\rm C}$ and $x$ for Na$_x$CoO$_2$.

The crystal structure of Na$_{0.9}$CoO$_2$ ($\alpha$-phase) is 
different from that of Na$_x$CoO$_2$ with $x \leq 0.75$ ($\gamma$-phase),
as seen in Fig.~\ref{fig:Phase}. 
That is, in the $\alpha$-phase, the sequence of the stacking of oxygen layers 
is represented as A-B-C-A-B-C, while in the $\gamma$ phase A-B-B-A,
where A, B, and C are named as similar for the stacking sequence 
in the face centered cubic close-packed structure.
\cite{NCO_structure_1} 
Also, Na ions occupy the octahedral sites in the  $\alpha$-phase, 
while the prism sites in the $\gamma$ phase.
\cite{NCO_structure_1} 
It is worth noting that the $\mu^+$ sites are bound to the O ions,  
indicating that the $\mu^+$ mainly feel the magnetic field 
in the [CoO$_2$] plane. 
Thus, the {\sf IC-SDW} is most unlikely to be caused 
by the misfit between the two subsystems, 
but to be an intrinsic behavior of the [CoO$_2$] plane. 

The end member of Na$_x$CoO$_2$, that is, 
a fully occupied Na$_1$CoO$_2$ phase can not be prepared 
by a conventional solid state reaction technique 
and/or a flux technique, \cite{NCO_structure_1, NCO_Mika2} 
while the related compound LiCoO$_2$ 
can be easily obtained. 
The structure of LiCoO$_2$ is isomorphous 
with $\alpha$-NaFeO$_2$ ($R\bar{3}m$) \cite{LCO_1}
and almost the same structure as that of 
the $\gamma$-Na$_x$CoO$_2$ phase,
if we ignore the difference of the occupancy 
between the Na and the Li plane. 
LiCoO$_2$ is reported to be diamagnetic down to 4.2~K.\cite{LCO_2}
Since the Co$^{3+}$ ions in LiCoO$_2$ are 
in a low-spin state $t_{2g}^6$ 
as in the case for Na$_x$CoO$_2$,
the diamagnetic behavior is expected. 
Therefore, NaCoO$_2$ is also speculated to 
lack magnetically ordered states. 

The occupancy of Co$^{4+}$ spins ($S$=1/2) 
on the two-dimensional triangular lattice {\sf 2DTL}
increases with decreasing $x$, 
so that the other end member, 
Na$_0$CoO$_2$ would have a half filled {\sf 2DTL}.
In other words, every lattice site is 
occupied by $S$=1/2.
If we employ the Hubbard model 
within a mean field approximation
as the basis for explaining the magnetism of such system;
\cite{MHonTL_1,MHonTL_2,MHonTL_3}
\begin{eqnarray}
 H&=&-t\sum_{<ij>\sigma}c_{i\sigma}^{\dagger}c_{j\sigma} + 
 U\sum_i n_{i\uparrow}n_{i\downarrow} ,
\label{eq:Hubbard}
\end{eqnarray}
where $c_{i\sigma}^{\dagger}(c_{j\sigma})$ 
creates (destroys) an electron with spin $\sigma$ on site $i$, 
$n_{i\sigma}=c_{i\sigma}^{\dagger}c_{i\sigma}$ 
is the number operator, 
$t$ is the nearest-neighbor hopping amplitude and 
$U$ is the Hubbard on-site repulsion.
The electron filling $n$ is defined as 
$n$ = (1/2$N$)$\sum_i^N n_i$,
where $N$ is the total number of sites.  

At $T$=0 and $n$=0.5 ({\it i.e.}, 
Na$_0$CoO$_2$),
as $U$ increases from 0, the system is first 
a paramagnetic metal up to $U/t \sim 3.97$, 
then changes into a metal 
with a spiral {\sf IC-SDW}, 
and then, at $U/t \sim 5.27$, 
a first-order metal-insulator transition occurs.\cite{MHonTL_1}
The lack of magnetic transitions for Na$_x$CoO$_2$ 
with $x$=0.6 and 0.65 suggests that $U/t\leq$3.97. 
This means that Na$_x$CoO$_2$ is unlikely to be 
a strongly correlated electron system, 
because $U\gg t$ for such system. 
This conclusion is in good agreement 
with the magnitude of $\gamma$ determined 
by the present $C_{\rm p}$ measurement 
($\gamma$=3.62$\pm$0.04~mJK$^{-2}$ per mol atom 
= 13.03$\pm$0.14~mJK$^{-2}$ per mol Co for
Na$_{0.6}$CoO$_2$). 
This value is 18.5 times larger than that of Cu.

Also the calculations predict that,
\cite{MHonTL_2, MHonTL_3}
as $n$ increases from 0, 
the magnitude of $U/t$ at the boundary 
between the paramagnetic and {\sf SDW} phases 
decreases with increasing slope 
($d(U/t)/dn$) up to $n$=0.75. 
Even for $U/t$=0, the {\sf SDW} phase is stable 
at $n$=0.75.
$U/t$ increases with further increasing $n$,
with decreasing slope. 
Therefore, the dome-shaped relationship 
found in the phase diagram 
(Fig.~~\ref{fig:Phase}) is 
qualitatively explained by the model calculations, 
although the maximum located at around $x$=0.8 
({\it i.e.} $n$=0.9). 
Furthermore, it is most likely that 
Na$_{0.75}$CoO$_2$ enters the {\sf C-SDW} state 
below 22~K rather than the ferrimagnetic state, 
because the magnetism in Na$_x$CoO$_2$ 
seems to be totally understood 
by Eq.~(\ref{eq:Hubbard}).

\subsection{\label{ssec:MM} magnitude of internal magnetic field}
If the {\sf SDW} state is induced by the competition 
between $U/t$ and $n$ in the CoO$_2$ {\sf 2DTL}, 
the nature of the {\sf SDW} state 
is considered to be essentially same 
in all the cobaltites. 
However, the internal magnetic field of the ordered state, 
$\nu_{\mu}$(0~K), is found to be $\sim$3~MHz for Na$_x$CoO$_2$,
while it is $\sim$60~MHz for 
[Ca$_2$CoO$_3$]$_{0.62}^{\rm RS}$[CoO$_2$] and
[Ca$_2$Co$_{4/3}$Cu$_{2/3}$O$_4$]$_{0.62}^{\rm RS}$[CoO$_2$].
 \cite{jun_PRB1,jun_PRB3} 
The muon locates
probably $\sim$0.1~nm away from the oxygen ions, 
and there is no space for it 
in the CoO$_6$ octahedra in the [CoO$_2$] plane 
as in the case for the high-$T_{\rm c}$ cuprates\cite{ICSDW_3}. 
This discrepancy is difficult to explain 
only by differences in the $\mu^+$ site 
experiencing the {\sf SDW} field. 
That is, even if the $\mu^+$s 
in Na$_x$CoO$_2$
locate in the vacant sites in the Na plane  
and those in the latter two cobaltites are 
bound to oxygen in the CoO$_2$ plane, 
the ratio between the bond length $d$ of Na-Co and O-Co 
is about 1.65.
Assuming that $\nu_{\mu}$(0~K)=60~MHz 
at the oxygen site in the CoO$_2$ plane,
$\nu_{\mu}$(0~K) at the Na site would be roughly 
$\sim$13~MHz, 
because the dipolar field is proportional to $d^{-3}$.
This is still four times larger than the experimental result. 

Hence, there should be the other reasons 
for the lower $\nu_{\mu}$(0~K) found in Na$_x$CoO$_2$. 
Considering the crystal structure of these cobaltites,
the distance between adjacent CoO$_2$ planes 
in Na$_x$CoO$_2$ is significantly smaller than those in 
[Ca$_2$CoO$_3$]$_{0.62}^{\rm RS}$[CoO$_2$] and
[Ca$_2$Co$_{4/3}$Cu$_{2/3}$O$_4$]$_{0.62}^{\rm RS}$[CoO$_2$]. 
Thus, the interlayer interaction 
between the CoO$_2$ planes in Na$_x$CoO$_2$ is considerably larger 
than in the other layered cobaltites. 
Such interaction is thought to weaken 
the two dimensionality of the CoO$_2$ plane. 
As a result, the magnitude of $\nu_{\mu}$(0~K) 
in Na$_x$CoO$_2$ may be smaller 
than those in the other layered cobaltites. 
Indeed, large transition widths, $\Delta T \sim 70$~K, 
were observed in
[Ca$_2$CoO$_3$]$_{0.62}^{\rm RS}$[CoO$_2$] and
[Ca$_2$Co$_{4/3}$Cu$_{2/3}$O$_4$]$_{0.62}^{\rm RS}$[CoO$_2$], 
while $\Delta T \sim 0$~K for 
Na$_x$CoO$_2$. 
This suggests an increase in spin frustration, 
reflecting the decrease in the two-dimensionality 
in Na$_x$CoO$_2$.
 
\section{\label{sec:Summary}Summary}
In order to elucidate the magnetism 
in 'good' thermoelectric layered cobaltites, 
$\mu^+$SR spectroscopy and 
heat capacity ($C_{\rm p}$) measurements 
were performed on single crystals of 
Na$_x$CoO$_2$ with $x$ = 0.6 and 0.9 
in the temperature range between 250 and 1.8~K. 

Both $\chi_c$, $C_{\rm p}$ and 
weak-transverse-field (wTF-) $\mu^+$SR 
measurements on Na$_{0.9}$CoO$_2$ indicated 
the existence of a magnetic transition at 19~K,
although no transitions were detected in Na$_{0.6}$CoO$_2$ 
down to 1.9~K. 
A clear oscillation in the zero-field (ZF-) $\mu^+$SR spectra,
fitted by a Bessel function, 
suggested that Na$_{0.9}$CoO$_2$ enters 
an incommensurate spin density wave state below 19~K
(=$T_{\sf SDW}$).
In addition, the {\sf IC-SDW} was found to propagate 
in the $c$ plane ({\it i.e.}, the CoO$_2$ plane), 
with oscillating moments directed along the $c$-axis,
similar to 
[Ca$_2$CoO$_3$]$_{0.62}^{\rm RS}$[CoO$_2$].

By reference to the previous $\mu^+$SR results 
on polycrystalline samples, 
a tentative magnetic phase diagram  
was obtained for Na$_x$CoO$_2$ with $x\geq$0.6.
The relationship between $T_{\sf SDW}$ and $x$ 
changed dome-shaped,
as well as the change in 
the high-$T_{\rm c}$ cuprates. 
Since this relationship was explained 
using the Hubbard model 
within a mean field approximation 
for two-dimensional triangle lattice 
of the CoO$_2$ plane, 
which is a common structural component 
for the all known thermoelectric layered cobaltites,
this bell-shape relationship is concluded to be 
a common behavior for the layered cobaltites.

The absence of the {\sf SDW} state in Na$_{0.6}$CoO$_2$ provided 
the upper limit for the magnitude of $U/t$ as 3.97, 
where $U$ denotes the Hubbard on-site repulsion and 
$t$ the nearest-neighbor hopping amplitude. 
This indicated that Na$_{0.6}$CoO$_2$ is
a moderately correlated electron system, 
although it is believed to be a strongly correlated system.
This conclusion was consistent with 
the measured electronic specific heat parameter ($\gamma$) of 
Na$_{0.6}$CoO$_2$, 
because $\gamma$ was almost equivalent to the calculated value 
without an electron correlation.

\begin{acknowledgments}
We thank Dr. S.R. Kreitzman and Dr. D.J. Arseneau of TRIUMF
for help with the $\mu^+$SR experiments. 
Also, we thank Dr. Y. Seno and Dr. C. Xia of Toyota CRDL, 
Mr. A. Izadi-Najafabadi and Mr. S.D. LaRoy 
of University of British Columbia 
for help with the experiments. 
We appreciate useful discussions with 
Dr. T. Tani and 
Dr. R. Asahi of Toyota CRDL, 
Professor U. Mizutani, Professor H. Ikuta and 
Professor T. Takeuchi of Nagoya University
and Professor K. Machida of Okayama University.  
This work was supported 
at UBC by the Canadian Institute for Advanced Research, 
the Natural Sciences and 
Engineering Research Council of Canada, 
and at TRIUMF by the National Research Council of Canada.  
\end{acknowledgments}

%\bibliography{apssamp}% Produces the bibliography via BibTeX.

\end{document}